# On performance of PBFT for IoT-applications with constrained devices

Yaroslav Meshcheryakov[1], Anna Melman[1], Oleg Evsutin[1], Vladimir Morozov[2], Yevgeni Koucheryavy[3], Senior Member, IEEE
[1]National Research University Higher School of Economics, 123458 Moscow, Russia
[2]Tomsk State University of Control Systems and Radioelectronics, 634050 Tomsk, Russia
[3]Tampere University, 33720 Tampere, Finland

Corresponding author: Oleg Evsutin (e-mail: evsutin.oo@gmail.com).

**ABSTRACT** Cyber-physical systems and the Internet of things (IoT) is becoming an integral part of the digital society. The use of IoT services improves human life in many ways. Protection against cyber threats is an important aspect of the functioning of IoT devices. Malicious activities lead to confidential data leakages and incorrect performance of devices are becoming critical. Therefore, development of effective solutions that can protect both IoT devices data and data exchange networks turns in to a real challenge. This study provides a critical analysis of the feasibility of using blockchain technology to protect constrained IoT devices data, justifies the choice of Practical Byzantine Fault Tolerance (PBFT) consensus algorithm for implementation on such devices, and simulates the main distributed ledger scenarios using PBFT. The simulation results demonstrate the efficiency of the blockchain technology for constrained devices and make it possible to evaluate the applicability limits of the chosen consensus algorithm.

## I. INTRODUCTION

Smart devices are part and parcel of the Internet of Things (IoT) in today's world, occupying an important place in people's lives. The rapid development of digital electronics, sensors, and communication systems has made it possible to create multifunctional smart devices that are propelling humanity towards a qualitative transition from the era of industrial progress to a new cybernetic era of development. People use a range of devices capable of storing and processing information and ready to integrate and interact with other devices. The functioning and integration of different IoT devices requires the development of large distributed systems for information transmission and storage. Distributed networks are necessary for data protection, collecting a large amount of data from a variety of decentralized sources and transferring them to data centers for processing [1-3].

Blockchain technology is an effective solution to the problem of collecting, transferring, storing, and protecting IoT devices data. This technology has received more attention in recent years [4-6], due to its properties of immutability and decentralization. Currently, blockchain technology has been researched and applied in various fields, such as the financial sector [7], e-health [8], access control [9], the Internet of vehicles (IoV) [10], the industrial IoT [11], and many others.

However, the development of high-performance blockchain solutions for IoT devices is challenging. The main distinguishing feature of IoT devices as opposed to other areas is the limitations on computing resources and the amount of memory, as well as strict requirements for energy saving. This imposes limits on the available computational complexity, which makes it difficult to apply classical consensus algorithms.

Based on the foregoing, this work analyses the blockchain technology applicability for constrained IoT devices and search for an effective consensus algorithm for them.

In this paper we provide the following contributions:
- We review applied blockchain solutions based on different consensus algorithms and argue the effectiveness of the PBFT consensus algorithm for its implementation on constrained IoT devices.
- We simulate key distributed ledger scenarios using the PBFT and evaluate the blockchain system performance.

- Based on the simulation results, we discuss the limits of applicability of the PBFT when developing a distributed ledger on constrained IoT devices.

The remainder of this paper is organized as follows. Section 2 describes blockchain technology and the main consensus algorithms. Section 3 justifies the choice of the PBFT consensus algorithm to study its applicability in IoT device networks. Section 4 describes research methods and presents simulation details. Section 5 contains the results of experiments and their discussion. The Conclusion summarizes, indicating the directions for future research.

## II. RELATED WORK

### A. BLOCKCHAIN

Blockchain technology was created in 2008 by Satoshi Nakamoto [12]. Blockchain is a decentralized technology that allows us to ensure the integrity of a distributed database that stores information about all transactions of system participants, without the participation of a trusted center. Transactions refer to some actions from a certain list, performed on tangible or intangible assets owned by users of the system. The transactions information is combined into blocks, which are chained together through hashing. Fig. 1 illustrates the general architecture of blockchain systems.

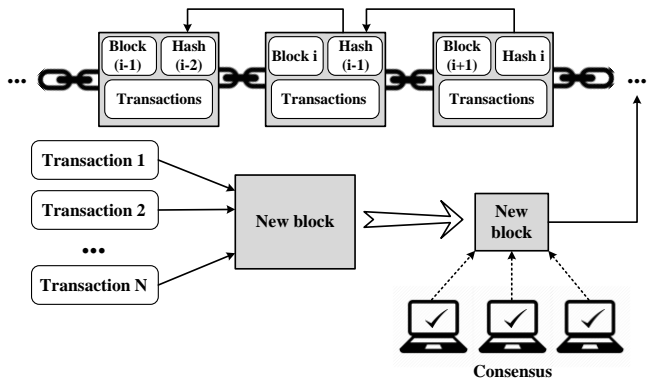

**FIGURE 1. The general architecture of blockchain systems.**

A special algorithm called the consensus algorithm is used to distribute identical copies of blocks between all system participants. It aims to make compromising blockchains a difficult task for a potential attacker [13]. The consensus algorithm is a mechanism for reaching agreement between network users, each of whom is interested only in personal gain and has no reason to trust all other participants [14, 15]. In other words, it is the way which the network nodes use to reach agreement on the order and composition of the stored data about the transactions they make. The consensus algorithm is a crucial part of a blockchain technology. It defines the architecture of the entire system and the order of interaction of network nodes. Different consensus algorithms can vary significantly in computational complexity and hardware requirements.

Recently, many authors have been actively investigating the blockchain technology implementation in a variety of areas, such as telemedicine [16], energy [17], supply chain tracking [18], and forensics [19]. The effectiveness of blockchain technology application in a specific area directly depends on the chosen consensus mechanism. Therefore, the study of the applicability of blockchain technology to a specific applied problem actually means choosing the most appropriate consensus algorithm and evaluating its performance indicators.

The following is an overview of studies in which blockchain technology is applied to solve data protection problems in different areas. These works are grouped in accordance with the consensus algorithms used in blockchain systems.

### B. PROOF OF WORK

The Proof of Work (PoW) consensus algorithm is one of the first algorithms to reach agreement in blockchain system [20]. The nodes of the PoW blockchain network compete for the right to add a new block to the ledger. To do this, they solve hard computational problems, and the solution correctness is easily verifiable. This protects the ledger from attempts to modify it. However, the disadvantage is excessive energy consumption. Blockchain developers in various application areas resort to the PoW algorithm.

In [21], a blockchain based on the PoW consensus algorithm is used in a decentralized system for government tenders. The proposed system architecture allows for control access to network nodes data based on authentication, ensuring transparency and security of the computing infrastructure for the implementation of government schemes and policies. The study [22] presents a blockchain protocol for IoV, using a dynamic PoW consensus algorithm. The protocol combines the use of smart contracts and physical unclonable functions to ensure trust in an IoV environment. The article [23] proposes a decentralized patient authentication system in a distributed network of hospitals using a blockchain. To reduce the authentication time and computational load, PoW and the addition of new blocks to the blockchain are performed only on devices that do not have significant energy and power limitations. The paper [24] describes the blockchain solution for providing the supply chain traceability for the textile and clothing industry using smart contract technology. The proposed solution is demonstrated by the example of traceability of organic cotton where PoW acts as the implemented consensus algorithm.

### C. PROOF OF STAKE

The Proof of Stake (PoS) consensus algorithm is the most common alternative to the PoW algorithm [25]. PoS was developed to overcome some of the challenges faced by PoW. First of all, this refers to reducing energy consumption, which is excessive for PoW. The main idea of PoS is a pseudo-random selection of nodes to generate the next block, depending on the share of units (tokens) owned by the node, as well as on its activity. The computational power of a node does not affect the ability of this node to generate a block.

The authors of [26] investigate the effectiveness of PoS for blockchain-based intrusion detection systems and propose several modifications. The study [27] proposes a validation control mechanism for Vehicular Ad-Hoc Networks (VANETs) using PoS. It allows us to decide whether to perform validation locally or move it to the edge or cloud infrastructure. In [28], the authors present a security data collection system for Mobile Ad Hoc Networks. A feature of this work is the focus on providing incentives for all participating nodes. The authors of [29] propose a blockchain system model based on Ethereum with a PoS consensus mechanism to ensure secure communication between drones and users for collecting and transmitting data in the Internet of Drones environment.

### D. DELEGATED PROOF OF STAKE

The Delegated Proof of Stake (DPoS) consensus algorithm further develops PoS ideas and at the same time is more efficient. The basic principle of DPoS is to select nodes for generating new blocks and maintain a distributed ledger through voting. These nodes are called delegates. The rank of each candidate to become delegate is determined based on the number of units owned by the nodes that voted for it. A list of delegates is formed from the nodes for which the largest number of participants voted. Delegates generate blocks for some time. After that, the vote is repeated and the list of delegates is updated [30].

This algorithm appeared later than PoW and PoS, so fewer DPoS-based blockchain systems have been developed and researched by now. Studies [31, 32] can be noted as examples. The paper [31] describes a blockchain-based solution for transparent and secure maintenance of a digital register of land assets. The authors of [32] present a low-latency secure authentication model for drones in smart cities. They use modified DPoS for drones as a consensus algorithm that does not require re-authentication and can decrease the number of nodes for the authentication process.

### E. PRACTICAL BYZANTINE FAULT TOLERANCE

The Practical Byzantine Fault Tolerance (PBFT) consensus algorithm [33] was originally developed as a mechanism to ensure the integrity of a distributed network. According to this algorithm, all nodes must participate in the voting process to add the next block. A two-thirds majority is required to reach consensus. PBFT algorithm exchanges messages between nodes quite intensively to ensure the network integrity. A lot of PBFT modifications have also been developed so far. For example, Delegated BFT differs from PBFT in that not all nodes participate in the voting process, but only some delegates. Simplified BFT is a Byzantine fault tolerance algorithm in which one validator creates and proposes a new block of transactions.

PBFT is a common solution for integrating blockchain technology and VANETs. For example, [34] proposes a blockchain architecture aimed at combating vulnerabilities to the so-called illusion attacks associated with false messages. For this, an intelligent selection of nodes participating in the consensus process is used. In [35], the authors propose a Proof of Driving protocol using PBFT, to randomize the selection of honest miners for the efficient generation of the blocks for blockchain-based VANET applications. Several more examples of such studies are presented in [36, 37]. In [38], it is proposed to use the concurrent PBFT consensus mechanism to solve the problem of the fast node expansion in the supply chain. At the same time, the authors propose to classify peers in the supply chain into several clusters using transaction analysis. Studies [39, 40] use the PBFT consensus algorithm in blockchain-based audit systems. The authors of [40] pointed out that the choice of the consensus algorithm affects the security of the proposed system. In their study, they use PBFT due to its high throughput and low latency.

### F. OTHER CONSENSUS ALGORITHMS

We have analyzed the most common consensus algorithms in practice. A detailed overview of existing consensus algorithms is beyond the scope of this study. However, it should be noted that there is a large number of such algorithms, and new ones are being actively developed.

For example, paper [41] presents a blockchain-based distributed carbon Emission Trading System. The authors propose the Delegated Proof of Reputation consensus mechanism. It takes into account the reputation of the participants in the system, which is determined by their contribution to reducing carbon emissions. Study [42] offers an original solution for balancing customer flow in shopping mall scenarios without expensive floor plan changes. The authors propose a blockchain-based diversion model for which they use a cascading consensus protocol inspired by MSig-BFT [43] and a mode of "execute-order-validate" in Hyperledger Fabric. In [44], the Proof of Virtual Voting consensus mechanism [45] is used for the blockchain-based crowdfunding platform. It assigns votes to developers based on mathematical calculations. The paper [46] proposes a blockchain architecture for industrial IoT devices based on a lightweight hash function and a synergistic multiple proof consensus mechanism. Another lightweight solution for the Industrial IoT is presented in [46]. The authors propose to use an energy-efficient Proof of Authentication consensus mechanism. Study [47] proposes a consensus scheme, dependent on WiFi technology, suitable for constrained devices.

Review papers detailing the state of the art in the development of consensus mechanisms can be found in [48-50].

## III. CHOOSING A CONSENSUS ALGORITHM FOR IOT DEVICES

We have demonstrated a variety of applications that use blockchain technology. Many of those have own specifics, which must be taken into account when developing a complex blockchain solution. As noted earlier, the choice of a consensus algorithm is an important task. Correct choice of consensus algorithm determines effectiveness of developed architecture in many cases.

A promising direction is development of secure fault-tolerant trusted cyber-physical systems based on the blockchain. An important challenge of development of such new systems is in its convenience and pace, since the benefit from their implementation directly depends on the time and financial cost. The main issues are in limited computational resources of IoT devices and the strict requirements for power consumption. The classic PoW algorithm is not suitable for solving this problem due to its high resource consumption and dependence on the mining equipment performance. PoS and DPoS security is based on the fact that the node has a certain number of units (tokens), so the node is interested in preserving and increasing this number. This interest guarantees the correct behavior of the node, but at the same time contributes to the possibility of network centralization. Some consensus algorithms are designed to solve a specific problem, so their use for this problem is highly effective. However, such consensus mechanisms are mostly not universal and cannot be practically used by a wide range of developers.

PBFT does not need high computing resources to reach consensus, so it can be used for constrained devices. PBFT has a fairly simple architecture that is easy to implement. This makes this consensus algorithm attractive for the rapid development of applied blockchain solutions. It is often used by various researchers when designing blockchain architectures for different tasks, but not all studies have a detailed justification for this choice. Also, it should be noted that the general research trend in this area is focused on the development of new consensus algorithms, and insufficient attention is paid to the study of the applicability of known algorithms in new areas. Therefore, in this paper, we present a study of the applicability of PBFT for IoT devices. The obtained results can be useful for researchers and developers of blockchain systems for IoT devices to assess the proposed solutions effectiveness.

## IV. RESEARCH METHODS AND SIMULATION DETAILS

Two main approaches can be used to evaluate the effectiveness of consensus algorithms and blockchain architectures based on them. The first option is to use the methods of mathematical statistics to analyze the data of a real blockchain system. This option is possible if the solution is implemented in the form of a large blockchain system that collects statistics of its work for a long time and provides it to researchers. The advantage of this approach is the ability to study real data, not modeled one. However, this approach is only suitable for a very limited range of blockchain systems such as cryptocurrency systems. In addition, when it comes to evaluating the blockchain solution applicability for specific application areas, it is obvious that there are usually no implementations available for research.

An alternative approach is a computational experiment. To implement this approach, a simulation environment is needed which provides specific conditions of the applied problem. To do this, it is necessary to implement a software simulator system with a common interface for the tested algorithms, or use ready-made software solution [51, 52]. The advantage of this approach is the ability to consider all the features of the blockchain system, as well as to trace in detail the behavior of various consensus algorithms in situations specific to this system. It should be noted that this approach is good for research purposes, but in practice, when developing new blockchain architectures, the use of this approach is associated with certain difficulties. A test environment design for simulating algorithms is a time-consuming and costly task, which complicates the transition to the direct development and implementation of a blockchain system. At the same time, the use of third-party software is not always possible, since often such simulators are designed to analyze a limited set of algorithms.

In this study, in order to evaluate the applicability of PBFT to IoT devices, we carry out a number of computational experiments. To do this, it is necessary to use software that most accurately simulates the real work of the blockchain technology. We have developed a simulator that exactly implements the original protocol described by the authors of [33].

To ensure the adequacy of the models under study, we have analyzed the characteristics of real constrained IoT devices in terms of computing power and data rate. Data rate is primarily determined by the Ethernet controller and the microcontroller unit (MCU). Available mass IoT devices primarily use the following basic hardware options:

1. AVR or STM8 series 8-bit MCU and Ethernet controller ENC28J60.

8-bit microcontrollers do not have a built-in Ethernet controller, so an external controller is required to provide network access. Such a combination of MCU and Ethernet controller allows sending data at rate of up to 10 Mbps and processing a small amount of information. This is mainly used to collect data from various sensors and process them in a simple way. The operating frequency of the MCU is up to 32 MHz for AVR and 24 MHz for STM8. If the ENC28J60 controller is replaced with a W5200 / W5500, the bandwidth and buffer size will increase, but there will not be a significant increase in data rate.

2. 32-bit microcontrollers (ST or NXP) with physical Ethernet interface.

The use of modern high-performance MCUs such as ST STM32 or NXP LPC series ones allows achieving high data rates up to 100 Mbps. The clock frequency varies quite widely, ranging from 72 MHz to 480 MHz, providing a performance of 61 - 1327 DMIPS.

3. ESP32 series systems-on-a-chip (SoC).

The ESP32 SoC contains low-bandwidth wireless communication interfaces such as Wi-Fi and Bluetooth. SoC clock frequency can range from 160 MHz to 240 MHz, and peak performance is 600 DMIPS, which is nearly equivalent to the IBM System / 370 model 158.

For the experimental studies, we focused on typical scenarios that usually disrupt IoT system performance:

- Increase in the number of nodes. As an IoT system functions and develops, the number of IoT devices in the network can increase. It is expected that the consensus algorithm provides stable operation in this case.
- Failure of nodes. When developing blockchain solutions for IoT systems, it is necessary to take into account the network nodes failing. The failure of the hardware of some devices must not lead to a malfunction of the whole system.
- Latency increase. The latency is associated with the characteristics of devices and network infrastructure, high network load, etc. It is important to investigate its effect on the system performance.

Thus, to evaluate the consensus algorithm effectiveness, it is necessary to investigate how different IoT scenarios affect the number of blocks that form a distributed ledger. The block size and the block generation rate are also important characteristics of the consensus algorithm.

Therefore, to comprehensively assess the applicability of PBFT for IoT devices, the following experiments were performed:

- study of the influence of block size on the system performance for a different number of nodes;
- assessment of the influence of the block generation rate on the consensus algorithm efficiency for a different number of nodes;
- study of the influence of the latency on the consensus algorithm efficiency for different latency distributions.

In the next section, we demonstrate and discuss the results of these experiments.

## V. EXPERIMENTAL RESULTS AND DISCUSSION

This section presents the results of simulating different scenarios of constrained devices network. We simulated the networks of different node numbers to evaluate the effectiveness of the PBFT consensus blockchain application for IoT devices.

The network operation was simulated for the same time (30 minutes) with recording of the number of committed blocks at a regular interval (every minute) for all experiments.

The structure of the data packet is described in Table I. The system uses a uniform data packet format for all types of messages. Depending on the type of messages, the nodes use the required fields and ignore the rest ones. For example, the hash value of a message within the protocol is 32 bytes, but if the message contains a transaction, the size of this field is 0. Sender ID and Recipient ID are fields of the lower layer protocol responsible for packet delivery. They are used only by the network manager, not by the nodes themselves. The Elliptic Curve Digital Signature Algorithm (ECDSA) is used to implement the electronic signature.

Fig. 2 shows the number of committed blocks per minute for a different block size. Each figure demonstrates experimental results for a network of 5, 10, 15, 20, and 25 nodes, respectively. The number of transactions in a block varies from 5 to 50.

The graphs demonstrate that increase in the number of transactions in a block reduces network performance. However, increase in the number of nodes in the network to 20 and more distorts this pattern. For a network of 20 nodes, more blocks of 10 transactions than blocks of 5 transactions were committed during the simulation. Better performance is observed when committing 20-30 transactions to a block than when committing 5-10 transactions to a block for the network of 25 nodes. The frequency of generating new blocks is inversely proportional to the block size if a small block size and a constant transaction generation period are used. Increasing in the frequency of block generation leads to increasing in the number of messages within the protocol, most of which are sent from each node to all others. As a result, the network becomes overloaded with packets, which reduces performance and leads to a decrease in the number of committed blocks. The experiment shows that the number of transactions in a block is a parameter that nonlinearly affects the performance of a network with a variable number of nodes.

TABLE I
THE STRUCTURE OF THE DATA PACKET

| Field name | Size, bytes |
|---|---|
| Sender ID | 4 |
| Recipient ID | 4 |
| Signature | 64 |
| Transaction | 1000 |
| Transaction type | 4 |
| Block | 8 |
| Time stamp | 4 |
| Node ID | 8 |
| Current view number | 8 |
| Message hash | 32 |
| Client request | 8 |
| Request number | 8 |

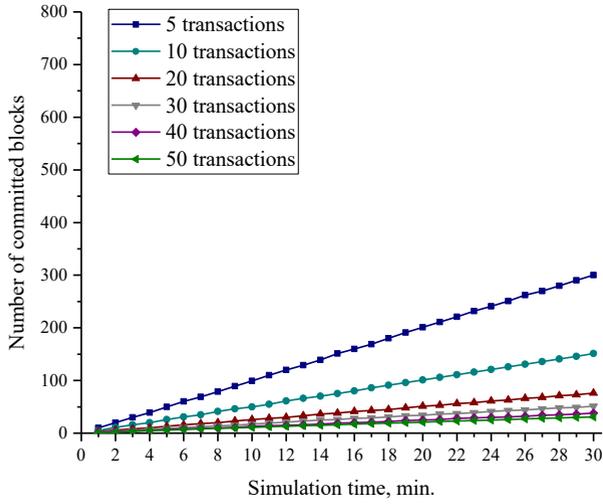

(a)

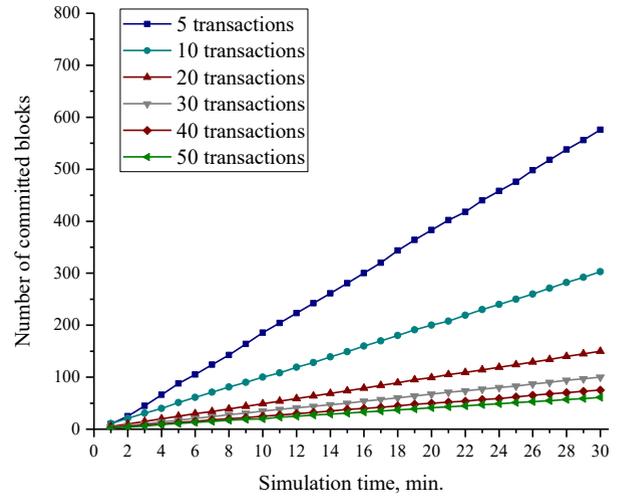

(b)

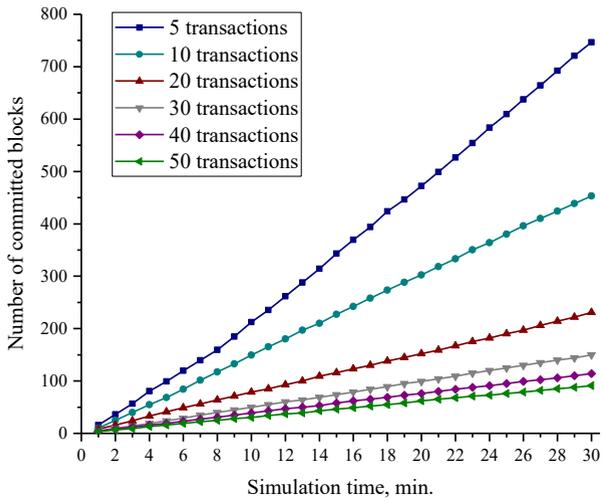

(c)

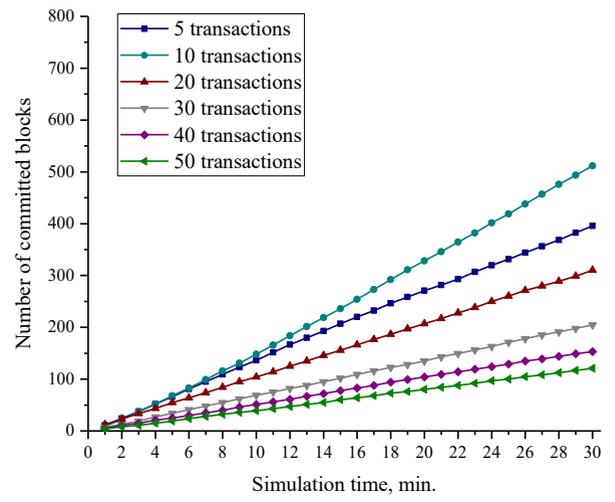

(d)

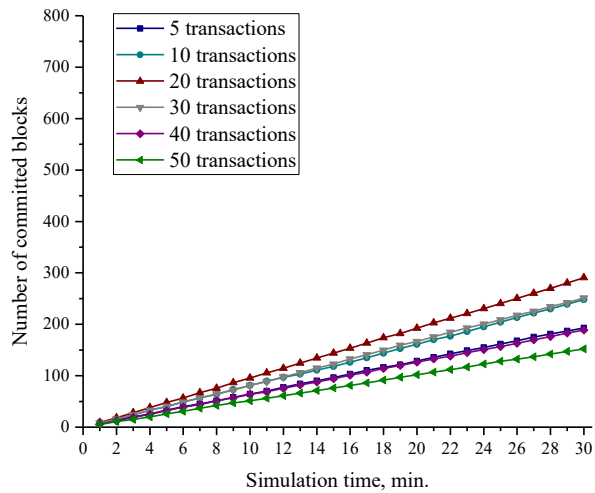

(e)

**FIGURE 2.** Block committed per minute with different size of blocks for: a) 5 nodes; b) 10 nodes; c) 15 nodes; d) 20 nodes; e) 25 nodes.

The results were obtained under simplified conditions that did not take into account the limitations on the device buffer size. Fig. 2a-2e illustrate the general trend. To assess the efficiency of the PBFT blockchain in the most realistic conditions, it is necessary to take into account that the hardware capabilities of the network interfaces of end devices are limited. This means that some of data may be lost during the operation of the network. The protocol under study provides for the transmission of repeated requests to commit the block in case of data loss. Increasing the number of requests decreases the share of the payload on the network. Therefore, the functionality of the network can be determined through the number of repeated requests for block committing. A large number of repeated requests indicates that new blocks are not being committed, and in fact the network is not functioning. Therefore, in order to assess the impact of the number of transactions in a block on system performance, we also examined how the block size affects the number of repeated requests to commit the block. We varied the number of nodes in the network and the number of transactions in the block, and measured the average number of repeated requests to commit the block sent by the nodes. The result of the experiment is illustrated in Fig. 3.

For clarity, the maximum number of repeated requests that can theoretically be achieved by one end device within the framework of this experiment is shown in Fig. 3 with a dotted line. Since the total simulation time is 30 minutes, and a retry request is sent once every 10 seconds, if necessary, the maximum number of retry requests is 180.

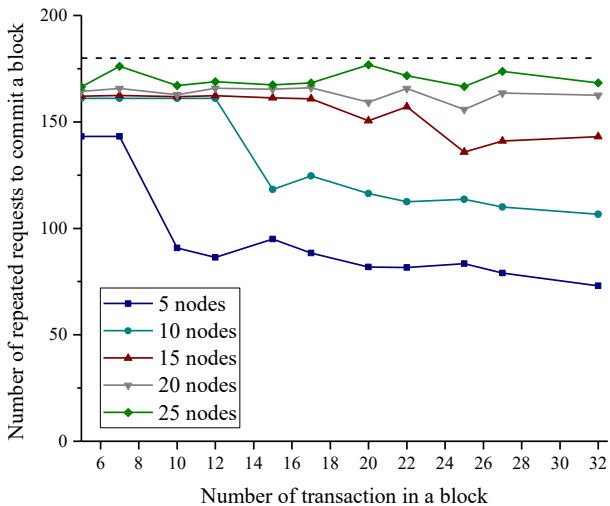

**FIGURE 3. The number of repeated requests to commit a block for different node numbers.**

The experimental results show that increasing the number of transactions in a block decreases the number of repeated requests. However, with an increase in the number of nodes, this decrease occurs with a larger block size. Implementing a larger block size reduces the number of blocks in the system and the load on the network, so each block is committed faster on average. With an increase in the number of nodes, the number of blocks increases, as a result the number of repeated requests also increases.

During the simulation, we evaluated the effect of the block generation rate on the network efficiency for a different number of nodes. We measured the number of committed blocks for different data generation period. The corresponding graphs are shown in Fig. 4. Each graph corresponds to one of the sizes of the simulated network.

The results of the experiment allow us to conclude that an increase in the data generation period leads to a decrease in the number of committed blocks if the network consists of a small number of nodes. The best result was achieved with the data generation period of 5 seconds on a network of 20 nodes (Fig. 4d). Under these conditions, the transactions generating rate allows nodes not to stand idle while waiting for a block to be committed, but it is not so high that the network is overloaded with packets. Increasing the network size to 25 nodes produces an effect similar to that seen in Fig. 2e and 2d. The line that previously corresponded to the highest number of committed blocks goes lower. Fig. 4e shows that the best performance is now observed for a period of 10 seconds instead of 5 seconds. This is due to the increase in network load caused by too intensive generation of transactions. Fig. 2 and 4 allow us to conclude that in a network with a large number of nodes, you need to choose a larger block size and data generation period in order to avoid network overload.

Delay in data transmission is common in practice. The delay can be caused by various reasons related to the characteristics of the devices and network configuration, or disruption to the network. In our study, we assessed the impact of latency on the performance of a blockchain with the PBFT consensus algorithm. We simulated incoming data packet processing latency. As part of the experiment, time delays were artificially introduced for processing an incoming message by a node. We considered uniform, normal, and exponential distributions of latency. After the node received the next message, a random number (delay time) was generated according to the selected distribution. After waiting for the generated amount of time, the node began processing the input.

The experimental results are shown in Fig. 5. Each figure shows a group of graphs of the dependence of the committed blocks number on time for different average delay times. The latency increasing negatively affects the efficiency of the blockchain. The worst result was demonstrated by the uniform distribution of latency, at which the least number of blocks was added to the blockchain. The best performance occurs with an exponential distribution of latency.

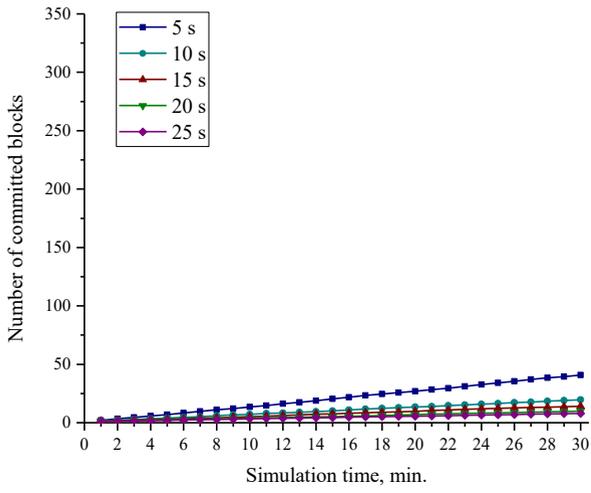

(a)

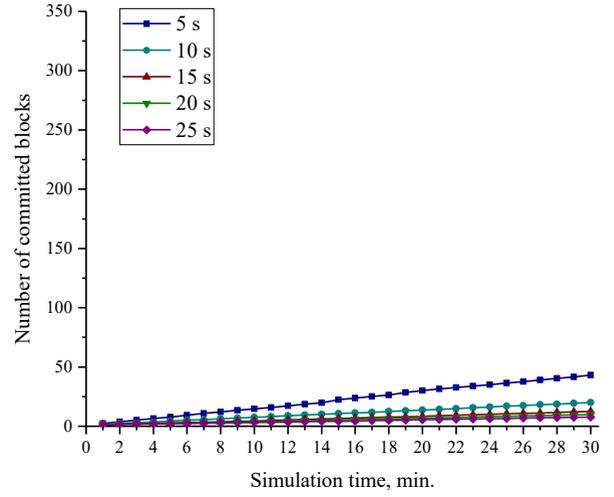

(b)

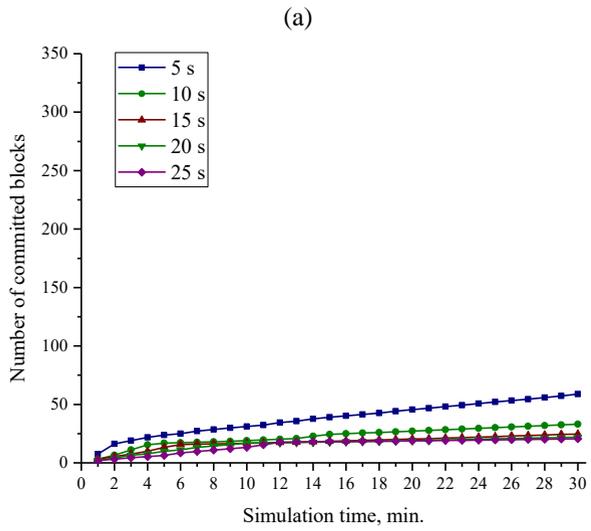

(c)

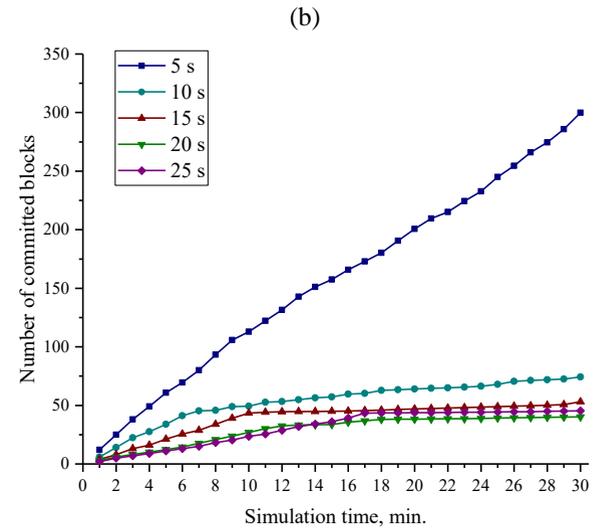

(d)

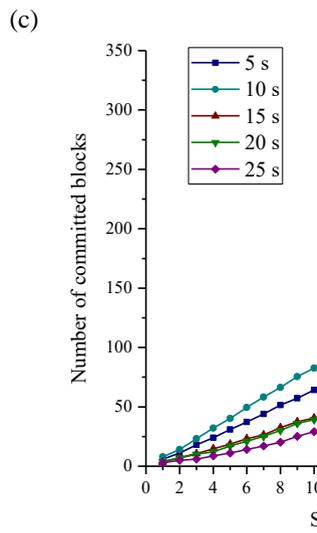

(e)

**FIGURE 4.  Block committed per minute with different data generation period for: a) 5 nodes; b) 10 nodes; c) 15 nodes; d) 20 nodes; e) 25 nodes.**

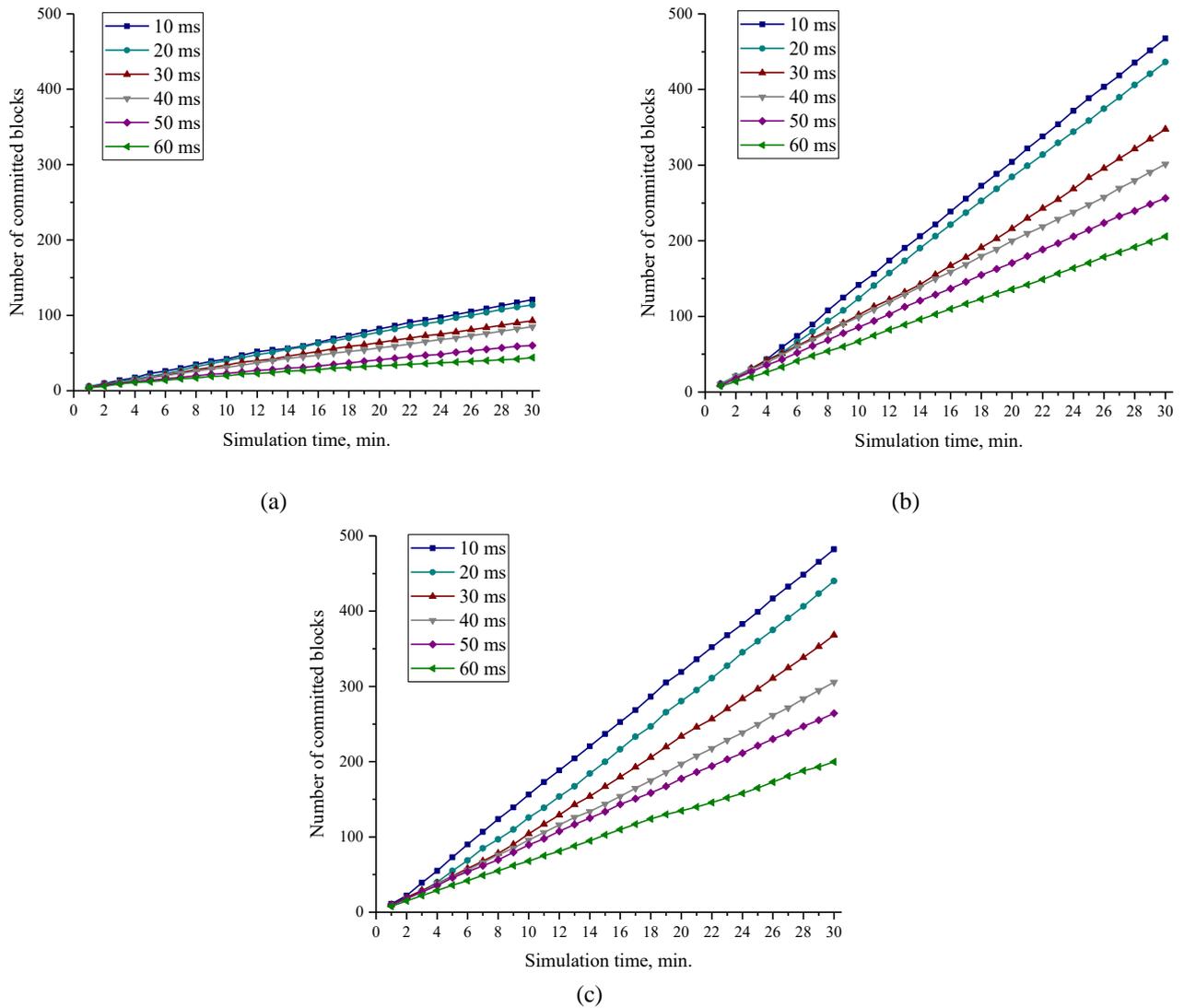

**FIGURE 5. Block committed per minute for a) uniform; b) normal; c) exponential latency distribution.**

Based on the results of the experiments, we can note the good performance of PBFT. The study of the dependence of the number of committed blocks on the block size showed that it is advisable to choose the block size consisting of 5-20 transactions. Another experiment, taking into account the limitation on the buffer size of the end devices, showed a similar result. The generation period, equal to 5-10 seconds, made it possible to add more blocks to the blockchain during the simulation than the longer generation period. It was also found that the number of committed blocks does not depend on the distribution of the delay time, but only on the average value of the delay. Increased latency negatively affects network performance.

Modern self-contained, battery-powered MCU-based devices are characterized by limitations in computing power, power consumption, memory, and bandwidth. This makes the development of distributed IoT systems challenging. The simulation results show that the PBFT algorithm performs well under these constraints. This makes it applicable for constrained device systems.

Based on the results of the network simulation, we also estimated the maximum number of devices with which PBFT can function. It is primarily determined by the throughput of the devices communication channel, the number of verification devices (nodes), and their computing power. The load from each node was recorded during the simulation. The load was about 85% with 25 operating nodes, and it was about 90% with 30 nodes. To simulate constrained devices like electronic implants, we reduced the packet length to 128 bits. This showed that the load was in the order of 45% with 30 devices. A further increase in the number of nodes led to an almost linear relationship. Interpolating the obtained dependencies, we can conclude that the maximum number of devices under these conditions is 70 units.

Low power consumption is one of the distinguishing features of PBFT. Considering this feature and the results of experiments, we provide examples of the most suitable applications for PBFT. They are united by the autonomy of the devices, as well as the small scale of the network. What they have in common are the self-contained devices and the small scale of the network.

1. Implantable medical devices.

Electronic implants are devices consisting of microelectronics, power supply and communication systems. These devices are implanted into the human body and operate for a long time in an autonomous way, which requires maximum energy savings . According to the principle of operation, implants are divided into permanently functioning and functioning for a short time, and spending the rest of the time in standby mode. The average operating time of an invasive implant is from 5 years to 14 years, the average operating time of a non-invasive implant is from 14 days to 3 months.

2. Self-contained telemedicine devices.

This group of devices includes such devices as plethysmographs, devices for determining a person's spatial orientation and systems for recording heart rate. These devices are characterized by autonomous operation, but the requirements for energy savings are less stringent than in the case of implanted devices. The task of these devices is to collect data within a given time. If the power supply is discharged, it can be recharged. The average life of such devices from a power source is from one hour to a month. The devices can function both in hospitals and outside. In a hospital environment, the device communicates directly with other devices and the hospital network. In the case of work outside the hospital, data exchange is carried out over a radio channel.

3. Self-contained devices for collecting and processing data for the IoT and the Industrial IoT

These devices operate in a pulsed autonomous mode for a long time, which usually ranges from 3 to 5 years. Most of the time the devices are in standby mode, data collection occurs at specified time intervals. The number of devices can vary greatly depending on the application. The distributed ledger system for data collection has a throughput in the range from 50 kbps to 10 Mbps. The transmission delay time is determined primarily by the MCU and the driver used.

Thus, the PBFT consensus algorithm is suitable for a wide range of IoT systems. Good performance combined with low resource consumption make this consensus algorithm applicable for constrained devices in various fields.

## VI. CONCLUSION

Currently, the use of IoT devices is becoming more widespread. The integration of IoT and blockchain technologies is a promising area. An important aspect of the blockchain schemes development is the choice of a consensus algorithm for a specific application area. In this study, we focus on the impact of the consensus algorithm on the performance of a blockchain system. We analyzed the applicability of the PBFT consensus algorithm to constrained IoT devices. To do this, we developed a simulation tool and performed a series of computational experiments to evaluate the effectiveness of PBFT in typical IoT scenarios. We investigated the dependence of the committed blocks number on the block size, data generation period and incoming data packet processing latency. Experiments show good PBFT performance for networks with a small number of nodes with a block size of 5-20 transactions and a data generation period of 5-10 seconds. Based on the simulation results, we estimated the maximum number of constrained devices in a PBFT-based blockchain system. This consensus algorithm allows for high performance for networks up to 70 nodes. This makes PBFT applicable to many types of IoT systems such as implanted medical devices, self-contained telemedicine devices, and small-scale systems of self-contained data collection and processing devices.

In our further work we plan to implement a network infrastructure using real constrained devices for a detailed study of the applicability of PBFT in this area in practice.

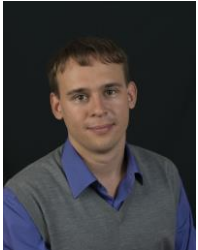

**YAROSLAV MESHCHERYAKOV** received the Specialist degree from Kuzbass State Technical University, Kemerovo, Russia, in 2014 and Candidate in Engineering degree from Tomsk State University, Tomsk, Russia, in 2018. He is currently an Associate Professor of the Department of Information Security of Cyber-Physical Systems at the Higher School of Economics, Moscow, Russia. His research interests include blockchain, Internet of Things, telecommunication technology, and e-healthcare.

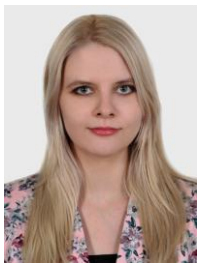

**ANNA MELMAN** graduated in information and analytical security systems from the Tomsk State University of Control Systems and Radioelectronics, Tomsk, Russia, in 2018. She is a junior research fellow of the Department of Information Security of Cyber-Physical Systems at the Higher School of Economics, Moscow, Russia. Her research interests include information security, digital steganography, and digital watermarking. She was awarded a medal of the Russian Academy of Sciences in 2017.

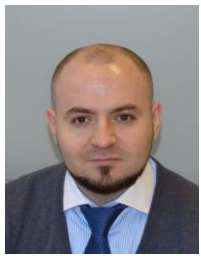

**OLEG EVSUTIN** received the Specialist degree from the Tomsk State University of Control Systems and Radioelectronics, Tomsk, Russia, and Candidate in Engineering degree from Tomsk State University, Tomsk, Russia, in 2009 and 2012, respectively. He is the Head of the Department of Cyber-Physical Systems Information Security, National Research University Higher School of Economics, Moscow, Russia. His current research interests include information security, steganography, digital watermarking, Internet of Things. He became the Laureate of Russian Federation Government Prize in Science and Technology for Young Scientists in 2018.

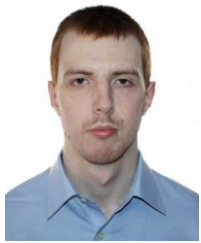

**VLADIMIR MOROZOV** is a student of Department of Complex Information Security of Computer Systems of Tomsk State University of Control Systems and Radioelectronics, Tomsk, Russia. His current research interests include blockchain technology, Internet of Things, and e-healthcare.

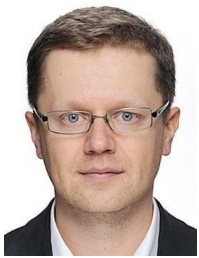

**YEVGENI KOUCHERYAVY** (SM'09) received the Ph.D. degree from the Tampere University of Technology, Finland, in 2004, where he is currently a Full Professor with the Unit of Electrical Engineering. He has authored numerous publications in the field of advanced wired and wireless networking and communications. His current research interests include various aspects in heterogeneous wireless communication networks and systems, Internet of Things and its standardization and nanocommunications.